\documentclass[12pt,reqno]{amsart}
\usepackage{amssymb,latexsym,amsmath}
\usepackage{amscd,amsfonts}
\usepackage{epsfig}

\newtheorem{theorem}{Theorem}
\newtheorem{lemma}{Lemma}

\newtheorem{corollary}{Corollary}

\theoremstyle{definition}
\newtheorem{remark}{Remark}

\numberwithin{equation}{section}

\begin{document}
\title{ On  the Helmholtz potential metric: the Isotherm Length-Work Theorem}
\author{Manuel Santoro}\address{Department of Mathematics and
Statistics, Portland State University, PO Box 751, Portland, OR
97207-0751, USA}
 \email{emanus@pdx.edu}
 Dedicated to: Francesco Santoro
 \qquad Benedetta Mazzuti
 \qquad Carlo Santoro
\begin{abstract}
In this paper we introduce the Isotherm Length-Work theorem using
the Helmholtz potential metric and the virial expansion of
pressure in inverse power of molar volume. The theorem tells us
what length of a thermodynamical system described by equation of
state through virial expansion along isotherms actually is with
such a metric. We also give explicit solutions for thermodynamic
length along isotherms in the case of first, second and third
order expansion.
\end{abstract}\bigskip
 \maketitle
\section{Introduction}
\par
J.W. Gibbs$^{6}$ introduced in his work a geometrical
interpretation of equilibrium thermodynamics which was followed by
C. Caratheodory$^{3,4}$ who proved that the state space is endowed
with the canonical contact structure that underlines the first law
of thermodynamics (energy balance law). Different representations
of this structure in a canonical (D'Arbois) chart are related to
different forms of the energy balance law written through internal
energy, entropy, Helmholtz free energy, etc.$^{8,11}$ \par R.
Hermann$^{7}$ and R. Mrugala$^{8}$ suggested that the "extended
phase space" of a homogeneous thermodynamic system endowed with
the contact structure does represent the natural geometrical space
for description of equilibrium thermodynamics.
\par
Later,a thermodynamic metric was explicitly introduced by F.
Weinhold$^{21}$ and, from a different point of view, by G.
Ruppeiner$^{12}$. These metrics were defined on the space of
thermodynamic states of a system using different thermodynamic
potentials, respectively internal energy U and entropy S, and set
of extensive variables, respectively $(S,V,N_{1},...)$ and
$(U,V,N_{1},...)$. It became clear, then, that a physical
interpretation of path-length between two states had to be
investigated. These studies were conducted using mostly the
internal energy formalism of a single component
system$^{16,18,19,20}$ and some results were obtained for systems
at constant entropy, volume, pressure and
temperature$^{16,18,19}$. In particular, it was shown$^{16}$ that,
using Weinhold metric for an Ideal gas, a reversible process at
constant temperature gives length equal to zero. Thus, the
following question arises: what is the meaning of thermodynamic
length of a quasi-static process along isotherms even for more
complex systems in which inter-particle interaction occurs?
\par
It is important to note that, in the molar energy representation,
the temperature T is function of the extensive variables s and v,
namely $T=T(s,v)$, and, therefore, analytical procedures become
considerably awkward. The standard approach, then, would be to
consider the Helmholtz free energy as thermodynamic potential
obtained by Legendre transformation of the energy function with
temperature and volume as independent variables and, then, define
a \textit{new} metric as the Hessian of the Helmholtz free energy.
Thus, this manuscript will unfold in a sequence of two main
points. First, we shall define and study the Helmholtz potential
metric of a two dimensional thermodynamic system. In particular,
we shall see that the tangent space at any point on the
equilibrium surface is the Lorentzian space $E^{1,1}$. Second, we
shall study path-length along isotherms using the virial expansion
in inverse power of molar volume.
\par
\mathstrut
\par
Let's, now, introduce the concepts of contact structure,
thermodynamic metric and thermodynamic length.
\par
We shall define the extended phase space as a $(2n+1)$-dimensional
manifold P endowed with the contact structure given by a
differential $1$-form $\theta$ such that$^{7,8}$
\par
\[
\theta\wedge{(d\theta)^{n}}\neq{0}
\]
\par
where $\theta$ is called the contact form.
\par
In a local (D'Arbois) chart $(\Phi,(Y_{i}, X_{i}))$ with
$i=1,...,n$, any contact form $\theta$ can be represented as$^{1}$
\par
\[
\theta=d\Phi-\sum^{n}_{1}Y_{i}dX^{i}
\]
\par
A Legendre manifold $S\subset{P}$ is a n-dimensional maximal
integral submanifold of the Pfaff equation $\theta=0$$^{1}$. On
such a manifold we consider $\Phi$ to be the thermodynamic
potential, $Y_{i}=\frac{\partial{\Phi}}{\partial{X^{i}}}$ to be
the intensive variables and $X_{i}$ to be the extensive variables
with $i=1,...,n$. Equilibrium states form such a maximal integral
surface of contact form $\theta$ in the space P by the choice of n
extensive variables and thermodynamic potential as function of
these variables. Equilibrium surface is then geometrically
described by constitutive relation $\Phi=\Phi(X_{i})$. Another
choice of external variables and thermodynamic potential leads to
another equilibrium surface corresponding, in general, to another
constitutive relation$^{8,11}$.
\par
On such equilibrium surface, thermodynamic metrics are defined by
the constitutive relation $\Phi=\Phi(X_{i})$. The general form is
given by
\par
\[
\eta_{\Phi}=\frac{\partial^{2}{\Phi}}{\partial{X^{i}}\partial{X^{j}}}dX_{i}\otimes{dX_{j}}
\]
\par
and the corresponding matrix representation is denoted by$^{10}$
\par
\[
\eta_{\Phi_{ij}}=\frac{\partial^{2}{\Phi}}{\partial{X^{i}}\partial{X^{j}}}
\]
\par
\mathstrut
\par
\textbf{Thermodynamic metrics.} Weinhold introduced a
metric$^{21}$ in the space of thermodynamic states as second
derivative of internal energy with respect to extensive variables
$X_{i}$ and $X_{j}$ , namely $\eta_{U_{ij}} =
\frac{\partial^{2}{U}}{\partial{X^{i}}\partial{X^{j}}}$ with $i,j
= 1,...,n$. In a general setting, constitutive relation $U
=U(X_{1}, ...,X_{n})$ represents the energy surface in which, for
example, $X_{1} = S$, $X_{2} = V$ ,etc., where S is the entropy
and V is the volume of our system. Such a metric gives us a way to
define distances and angles and, therefore, it enables us to study
the geometry of the energy surface.
\par
Ruppeiner, instead, introduced a metric$^{13}$ by the choice
$\Phi=S$ as thermodynamic potential and defined it as second
momenta of entropy with respect to the fluctuations, namely
$\eta_{S_{ij}} =-
\frac{\partial^{2}{S}}{\partial{X^{i}}\partial{X^{j}}}$. In this
case constitutive relation $S=S(X_{1}, ...,X_{n})$ represents the
entropy surface in which, for example, $X_{1} = U$, $X_{2} = V$
,etc.
\par
As we have already mentioned, a different choice of thermodynamic
potential and extensive variables leads to a different equilibrium
surface geometrically described by a certain constitutive
relation. Since the most familiar thermodynamic potentials are the
Legendre transformations$^{2}$ of the internal energy, namely
Helmholtz free energy, Enthalpy, Gibbs free energy, and since this
manuscript is concerned with the meaning of thermodynamic length
which it has been studied mostly using Weinhold
metric$^{14,15,16,18,19,20}$, we shall focus our attention to the
energy surface and its geometry. We shall see that Legendre
transformations of a thermodynamic potential change the "nature"
of the thermodynamic state space.
\par
\mathstrut
\par
\textbf{Thermodynamic length.} The metric $\eta_{U_{ij}}$ can be
interpreted locally as the distance$^{16}$ between the energy
surface and the linear space tangent to this surface at some point
where $\eta_{U_{ij}}$ is evaluated. Indeed, let's denote by
coordinates $(X^{0}_{1},...,X^{0}_{n})$ a particular energy state.
The tangent space is attached to the energy surface at point
$(U^{0},X^{0}_{1} ,...,X^{0}_{n})$. If we move away a little to a
new energy state $(X_{1},...,X_{n})$ then the
availability$^{14,16}$ or the available work$^{14}$ of the system
is the distance between the point on the surface
$(U,X_{1},...,X_{n})$ and the tangent space. This is naturally a
local interpretation since it requires just small displacements,
like for fluctuations, from the given point
$(U^{0},X^{0}_{1},...,X^{0}_{n})$ on the surface.
\par
On the other hand, we could study thermodynamic length taking the
metric $\eta_{U_{ij}}$ globally. In this situation we consider a
path $\phi$ on the energy surface between two states $a_{0}$ and
$a_{1}$ and study the length of the path
\par
\[
L_{a_{0}a_{1}}=\int^{a_{1}}_{a_{0}}{[\sum_{i,j}\eta_{U_{ij}}dX_{i}dX_{j}]^{\frac{1}{2}}}
\]
\par
It was shown$^{16}$ that the thermodynamic length L does, in
general, represent the change in mean molecular velocity depending
on the particular nature of the thermodynamic process defining the
path $\phi$ and that its dimension is square root of energy. But
thermodynamic length was explicitly studied$^{16}$ just in the
Ideal case. In particular, it was found that, for a reversible
adiabatic Ideal Gas from state $(p_{0},V_{0})$ to state
$(p_{1},V_{1})$, length represents the change in flow velocity of
a gas undergoing an isentropic expansion, like in rarefaction
waves, and it is given by$^{16}$
\par
\[
L^{s}=\frac{2}{\gamma-1}\sqrt{\gamma{p_{0}}V_{0}}[1-(\frac{p_{1}}{p_{0}})^{\frac{\gamma-1}{2\gamma}}]
\]
\par
with $\gamma=\frac{C_{p}}{C_{v}}$.
\par
Moreover, in our previous manuscript, we have shown an
explicit$^{20}$ relation between thermodynamic length and work for
an isentropic Ideal and quasi-Ideal Gas along isotherms, namely
\par
\[
L^{s}=\sqrt{\frac{1}{RT}}W
\]
\par
Such a relation was considered to be a way to measure the amount
of work done by the system along isotherms. But its interpretation
in relation with work turned out to be much more complex than what
we had expected. Indeed, we realized that such a case was the
trivial one, namely that length in an isentropic Ideal and
quasi-Ideal Gas along isotherms is zero, and gave a generalization
of that relation no longer at constant temperature$^{18}$. In
particular, we found that thermodynamic length of an isentropic
Ideal or quasi-Ideal Gas measures the difference of the square
roots of the energies of two given states, namely$^{18}$
\par
\[
L^{s}=2\sqrt{\frac{c_{p}}{R}}[\sqrt{u_{2}+W_{in}}-\sqrt{u_{2}}]=2\sqrt{\frac{c_{p}}{R}}[\sqrt{u_{1}}-\sqrt{u_{2}}]
\]
\par
and
\par
\[
-L^{s}=2\sqrt{\frac{c_{p}}{R}}[\sqrt{u_{2}}-\sqrt{u_{2}-W_{out}}]=2\sqrt{\frac{c_{p}}{R}}[\sqrt{u_{2}}-\sqrt{u_{1}}]
\]
\par
where $W_{in}$ and $W_{out}$ are the work done on the system and
the work done by the system. (Note that we require length to be
positive). Therefore thermodynamic length is zero if there is no
work.
\par
\mathstrut
\par
So far we have been able to physically interpret thermodynamic
length for an isentropic Ideal and quasi-Ideal Gas. As a special
case we have shown that along isotherms such a length vanishes.
Thus, the following question arises, namely: what is the physical
meaning of length for an isothermal thermodynamic system?
\par
\begin{remark}
Note that we are no longer considering thermodynamic systems at
constant entropy. Just constant temperature.
\end{remark}
\par
Naturally, we cannot use the same mathematical approach using
Weinhold metric on the equilibrium surface described by
constitutive relation $u=u(s,v)$. We would like to \textit{change}
set of extensive variables in such a way to include temperature as
one of them. The standard way to do that is to consider the
Legendre transformation of internal energy which replaces the
molar entropy with the temperature. Therefore, we would need to
consider the molar Helmholtz free energy with corresponding energy
surface described by constitutive relation $f=f(T,v)$ as the
\textit{natural} setting for such a problem.
\par
\section{Thermodynamic length with the Helmholtz potential metric}
\par
We have already mentioned that thermodynamic metrics are
geometrically defined on Legendre submanifolds (equilibrium
surfaces) of the thermodynamical phase space by the choice of a
set of extensive variables and of a thermodynamic potential like
internal energy, entropy, Helmholtz free energy, enthalpy, etc. By
such a choice we are able to study, through constitutive relation,
the geometrical structure of that particular equilibrium surface.
It is on such equilibrium surface that we can define a metric as
second derivative of the thermodynamic potential with respect to
the extensive variables. All these metrics are naturally
well-defined on the corresponding Legendre submanifold through
Legendre transformation.
\par
\mathstrut
\par Here we will consider the Helmholtz free energy as
thermodynamic potential and we will study the metric of a
two-dimensional thermodynamic system defined by such a choice. We
shall see that the equilibrium surface defined by constitutive
relation $f=f(T,v)$ has, as a tangent space at any point, the
Lorentzian space $E^{1,1}$ since the eigenvalues of the relative
matrix metric have opposite sign, as long as we avoid points of
degeneracy. The length of any vector on such a space is either
positive, zero or pure imaginary. Naturally, the length of any
curve (thermodynamic process) on the equilibrium surface can be
parametrized and, thus, can be expressed in terms of rate of
change of position vectors with respect to the parameter. Such
vectors belong to the Lorenztian space at any point. For paths in
a constant direction, length is either positive, zero or pure
imaginary. We will define a volume-like vector, a temperature-like
vector and a null vector on the Lorenztian space $E^{1,1}$ at each
point q on the surface S.
\par
We shall see that length is computed in its generality using
virial expansion of pressure in inverse molar volume and just in
the Ideal and quasi-Ideal case is proportional to work along
isotherms.
\par
\mathstrut
\par
It is known that the Helmholtz (molar) potential $f$ is the
Legendre transformation of the molar internal energy $u$ that
replaces the molar entropy $s$ by the temperature T as independent
variable. That is
\par
\[
f=f(T,v)
\]
\par
Now, since $f=u-Ts$, we have the following differential,$^{2}$
\par
\[
df=-sdT-pdv
\]
\par
with
\par
\[
(\frac{\partial f}{\partial T})_{v}=-s
\]
\par
and
\par
\[
(\frac{\partial f}{\partial v})_{T}=-p
\]
\par
If we define the metric
$\eta_{f_{ij}}=\frac{\partial^{2}{f(x)}}{\partial{x_{i}}\partial{x_{j}}}$,
we have
\par
\begin{equation}
\eta_{f_{ij}}=
\begin{pmatrix}
 \frac{-c_{v}}{T} & -\frac{\alpha}{k_{T}}\\
 -\frac{\alpha}{k_{T}} & \frac{1}{vk_{T}}
\end{pmatrix}
\end{equation}
\par
where
\par
\begin{enumerate}
\item $c_{v}$ is the molar heat capacity at constant volume:
\[
c_{v}=T(\frac{\partial s}{\partial T})_{v}\qquad,
\]
\item $c_{p}$ is the molar heat capacity at constant pressure:
\[
c_{p}=T(\frac{\partial s}{\partial T})_{p}\qquad,
\]
\item $\alpha$ is the thermal coefficient of expansion:
\[
\alpha=\frac{1}{v}(\frac{\partial v}{\partial T})_{p}\qquad,
\]
\item $\kappa_{T}$ is the isothermal compressibility:
\[
\kappa_{T}=-\frac{1}{v}(\frac{\partial v}{\partial p})_{T}\qquad.
\]
\end{enumerate}
\par
Local conditions of stability require that the Helmholtz free
energy be a concave function of the temperature and a convex
function of the volume$^{2}$. It is easy to see that
$\det{(\eta_{f_{ij}})}=-\frac{c_{p}}{Tv\kappa_{T}}=\frac{c_{p}}{T}(\frac{\partial{p}}{\partial{v}})_{T}$
and that the characteristic equation of $(2.1)$ is given by
\par
\begin{equation}
\lambda^{2}+(\frac{c_{v}}{T}-\frac{1}{v\kappa_{T}})\lambda-\frac{c_{p}}{Tv\kappa_{T}}=0
\end{equation}
\par
It follows that the eigenvalues are given by
\par
\begin{equation}
\lambda_{1/2}=\frac{1}{2}[(\frac{1}{v\kappa_{T}}-\frac{c_{v}}{T})\pm{\sqrt{\Delta}}]
\end{equation}
\par
where
\par
\begin{equation}
\Delta=(\frac{1}{v\kappa_{T}}+\frac{c_{v}}{T})^{2}+4(\frac{\alpha}{\kappa_{T}})^{2}>0
\end{equation}
\par
Now, since $\Delta$ is always positive, the eigenvalues
$\lambda_{1/2}$ are both real and distinct and, since
$\det{(\eta_{ij})}=\lambda_{1}\lambda_{2}$, then we have the
following result
\par
\begin{lemma}
Let $T>0$. Since $c_{p}-c_{v}=\frac{vT\alpha^{2}}{\kappa_{T}}$,
then
\par
If $c_{p}>0$ and $(\frac{\partial{p}}{\partial{v}})_{T}<0$ then
$\det{(\eta_{f_{ij}})}<0$and $\lambda_{1}<0$, $\lambda_{2}>0$.
\par
\end{lemma}
\par
Let's assume that the eigenvalues are both non-zero. Then the
metric $(2.1)$ is diagonalizable. Since $\eta_{f_{ij}}$ is a real
symmetric matrix, it can be diagonalized by an orthogonal change
of basis of the tangent space $T_{q}S$ at point q to the surface
S. In the eigenvector basis, the shape of the equilibrium surface
S becomes obvious. Direction along eigenvectors with negative
eigenvalues have curvature downward and direction with positive
eigenvalues have upward curvature. Moreover, each eigenvector
would represent a particular perturbation of the surface.
Naturally, both eigenvalues and eigenvector would depend on the
point $(T,v)$. It follows that the matrix of eigenvalues is given
by
\par
\begin{equation}
\Lambda_{ij}=\frac{1}{2}
\begin{pmatrix}
 (\frac{1}{v\kappa_{T}}-\frac{c_{v}}{T})-\sqrt{\Delta} & 0\\
 0 & (\frac{1}{v\kappa_{T}}-\frac{c_{v}}{T})+\sqrt{\Delta}
\end{pmatrix}
\end{equation}
\par
The corresponding eigenvectors corresponding to $\lambda_{1}$ and
$\lambda_{2}$ are given by
\par
\begin{equation}
\xi_{1}=
\begin{pmatrix}
 1 \\
 -\frac{\kappa_{T}}{2\alpha}[(\frac{1}{v\kappa_{T}}+\frac{c_{v}}{T})-\sqrt{\Delta}]
 \end{pmatrix}
\end{equation}
\par
and
\par
\begin{equation}
\xi_{2}=
\begin{pmatrix}
 \frac{\kappa_{T}}{2\alpha}[(\frac{1}{v\kappa_{T}}+\frac{c_{v}}{T})-\sqrt{\Delta}] \\
 1
\end{pmatrix}
\end{equation}
\par
Now, since the two eigenvalues are distinct then the set
$[\xi_{1},\xi_{2}]$ is linearly independent.
\par
Let's denote, now, the matrix $P=(\xi_{1},\xi_{2})$. In
particular,
\par
\begin{equation}
P=
\begin{pmatrix}
 1 & \frac{\kappa_{T}}{2\alpha}[(\frac{1}{v\kappa_{T}}+\frac{c_{v}}{T})-\sqrt{\Delta}] \\
 -\frac{\kappa_{T}}{2\alpha}[(\frac{1}{v\kappa_{T}}+\frac{c_{v}}{T})-\sqrt{\Delta}]
 & 1
\end{pmatrix}
\end{equation}
\par
and the inverse $P^{-1}$ is given by
\par
\begin{equation}
P^{-1}=\frac{1}{1+\frac{\kappa^{2}_{T}}{4\alpha^{2}}[(\frac{1}{v\kappa_{T}}+\frac{c_{v}}{T})-\sqrt{\Delta}]^{2}}
\begin{pmatrix}
 1 & -\frac{\kappa_{T}}{2\alpha}[(\frac{1}{v\kappa_{T}}+\frac{c_{v}}{T})-\sqrt{\Delta}] \\
 \frac{\kappa_{T}}{2\alpha}[(\frac{1}{v\kappa_{T}}+\frac{c_{v}}{T})-\sqrt{\Delta}]
 & 1
\end{pmatrix}
\end{equation}
\par
It is evident that $\eta_{f_{ij}}$ can be decomposed in the very
special form
\par
\begin{equation}
\eta_{f_{ij}}=P\Lambda_{ij}{P^{-1}}
\end{equation}
\par
where P is a matrix composed of eigenvectors, $P^{-1}$ is its
inverse and $\Lambda_{ij}$ is the matrix of eigenvalues of
$\eta_{f_{ij}}$.
\par
Let's, now, define with $E_{\lambda_{1}}$ and $E_{\lambda_{2}}$
the eigenspaces of $\lambda_{1}$ and $\lambda_{2}$. Naturally, the
basis for the one-dimensional eigenspaces $E_{\lambda_{1}}$ and
$E_{\lambda_{2}}$ are given by $\xi_{1}$ and $\xi_{2}$. \par If we
denote by $|\xi_{i}|$, $i=1,2$, the length of the eigenvectors,
then we can normalize them obtaining a orthonormal basis for the
two-dimensional tangent space at any point q on the surface
considering the important fact that normalizing a vector of
imaginary length can require multiplication by a negative scalar
\par
\begin{equation}
B=(\frac{\xi_{1}}{|\xi_{1}|},\frac{\xi_{2}}{|\xi_{2}|})
=(\xi^{-}_{1},\xi^{-}_{2})
\end{equation}
\par
The tangent space $T_{q}S$ at any point q on S is a vector space
endowed with a pseudo-Riemannian metric given by
\par
\begin{equation}
\Lambda_{ij}=\frac{1}{2}
\begin{pmatrix}
 (\frac{1}{v\kappa_{T}}-\frac{c_{v}}{T})-\sqrt{\Delta} & 0\\
 0 & (\frac{1}{v\kappa_{T}}-\frac{c_{v}}{T})+\sqrt{\Delta}
\end{pmatrix}
\end{equation}
\par
Such a metric at any point on the domain, say $(T_{0},v_{0})$, is
equivalent to a Lorentz metric of the form
\par
\begin{equation}
\Lambda_{ij}|_{(T_{0},v_{0})}=
\begin{pmatrix}
 -1 & 0\\
 0 & 1
\end{pmatrix}
\end{equation}
\par
The tangent space $T_{q}S$ at a point q on the surface S is,
therefore, a Lorentzian 2-space and denoted by $E^{1,1}$.
\par
This implies that the length of any vector is either positive,
zero or pure imaginary.
\par
Now, it is known that the equilibrium surface defined by the
energy function $u=u(s,v)$ has a metric at any point of its domain
of the form
\par
\[
\Lambda_{ij}|_{(s_{0},v_{0})}=
\begin{pmatrix}
 1 & 0\\
 0 & 1
\end{pmatrix}
\]
\par
Moreover, the Helmholtz potential $f$ is the Legendre
transformation of u which replaces the molar entropy $s$ by the
temperature $T=\frac{\partial{u}}{\partial{s}}$. Then, it is
exactly the one variable Legendre transformation which change the
signature of the metric. In other words, it locally change the
Euclidean metric in a Lorentzian one and viceversa.
\par
\begin{remark}
Note that the same argument is true in case we would consider
Enthalpy as thermodynamic potential which is the Legendre
transformation of internal energy that replaces the molar volume
by the pressure as independent variables. In our future work,
we'll show that also such a metric is Lorentzian. As we will show
that, given Gibbs free energy as thermodynamic potential, which is
a double-variable Legendre transformation, the metric related to
it has signature $(-1,-1)$ which is equivalent to the Euclidean
signature $(1,1)$. As far this paper is concerned we are just
considering length along isotherms and ,therefore, we leave such a
remark as introduction to future work.
\end{remark}
Let's, thus, define positive length to be volume-like and pure
imaginary to be temperature-like. We stress again that pure
imaginary length is temperature-like due to the one variable
Legendre transformation which replaces s with T.
\par
From now on, we'll study length along isotherms. An isothermal
process typically occurs when a system is in contact with an
outside thermal reservoir, and the system changes slowly enough to
allow it to adjust to the temperature of the reservoir.
\par
\mathstrut
\par
Now, considering
\par
\begin{equation}
L_{a_{0}a_{1}}=\int^{a_{1}}_{a_{0}}{[\sum_{i,j}\eta_{f_{ij}}dX_{i}dX_{j}]^{\frac{1}{2}}}
\end{equation}
\par
to be the length of a path between two states $a_{0}$ and $a_{1}$,
the thermodynamic length with the Helmholtz potential metric
becomes
\par
\begin{equation}
L=\int{[-\frac{c_{v}}{T}(dT)^{2}-2\frac{\alpha}{\kappa_{T}}dTdv+\frac{1}{v\kappa_{T}}(dv)^{2}]^{\frac{1}{2}}}
\end{equation}
\par
\begin{equation}
=\int^{\xi_{f}}_{\xi_{i}}[-\frac{c_{v}}{T}(\frac{dT}{d\xi})^{2}-2\frac{\alpha}{\kappa_{T}}\frac{dT}{d\xi}\frac{dv}{d\xi}+\frac{1}{v\kappa_{T}}(\frac{dv}{d\xi})^{2}]^\frac{1}{2}d\xi
\end{equation}
\par
Using $(2.12)$, we have that
\par
\begin{equation}
(dL)^{2}=\lambda_{1}(dT)^{2}+\lambda_{2}(dv)^{2}
\end{equation}
\par
The expression above is not positive definite. Therefore the usual
concept of length has to be abandoned.
\par
In particular,
as we stated previously, we consider constant directional paths in
which we also allow zero and pure imaginary length. As mention
above, we will restrict our attention to the study of
thermodynamic length at constant temperature which is given by
\par
\begin{equation}
L^{T}=\int{\sqrt{\frac{1}{v\kappa_{T}}}dv}=\int{\sqrt{(-\frac{\partial{p}}{\partial{v}})_{T}}dv}=\int{\sqrt{-\frac{T}{c_{p}}\det{\eta_{ij_{f}}}}dv}=\int{\sqrt{\eta_{22}}dv}
\end{equation}
\par
\section{The Isotherm Length-Work Theorem}
\par
The \textit{Isotherm Length-Work Theorem} uses the virial
expansion in inverse power of molar volume which is given
by,$^{2}$
\par
\begin{equation}
p=\frac{RT}{v}+\frac{RTB(T)}{v^{2}}+\frac{RTC(T)}{v^{3}}+\frac{RTD(T)}{v^{4}}+...
\end{equation}
\par
where $B(T)$, $C(T)$, etc. are the virial coefficients.
\par
If we expand p up to the n-th power, then, we might express
$(3.1)$ as
\par
\begin{equation}
p=\frac{RT}{v}+\frac{RTB(T)}{v^{2}}+\frac{RTC(T)}{v^{3}}+\frac{RTD(T)}{v^{4}}+...+\frac{RTY(T)}{v^{n-1}}+\frac{RTZ(T)}{v^{n}}
\end{equation}
\par
where $Y(T)$ and $Z(T)$ are the $(n-1)$-th and the $n$-th virial
coefficients.
\par
Now, since the temperature is constant, say $T=T_{0}$, let's set
$B(T_{0})=B$, $C(T_{0})=C$, etc. and so, recalling the second
integral in $(2.18)$, we have
\par
\[
L^{T}=\int{\sqrt{(-\frac{\partial{p}}{\partial{v}})_{T}}dv}
\]
\par
\begin{equation}
=\int{\sqrt{\frac{RT}{v^{2}}+\frac{2RTB}{v^{3}}+\frac{3RTC}{v^{4}}+...+\frac{(n-1)RTY}{v^{n}}+\frac{nRTZ}{v^{n+1}}}}dv
\end{equation}
\par
\begin{theorem}
\textbf{Isotherm Length-Work Theorem}.
\par
Let T and v be non-zero. Then, along isotherms, thermodynamic
length is given by any of the following:
\par
\[
L^{T}=\frac{1}{\sqrt{RT}}[n\int{\frac{pdv}{\sqrt{1+\frac{2B}{v}+\frac{3C}{v^{2}}+...}}}-(n-1)\int{\frac{RTdv}{v\sqrt{1+\frac{2B}{v}+\frac{3C}{v^{2}}+...}}}
\]
\par
\begin{equation}
-(n-2)\int{\frac{RTBdv}{v^{2}\sqrt{1+\frac{2B}{v}+\frac{3C}{v^{2}}+...}}}-...]
\end{equation}
\par
\[
=\frac{n}{\sqrt{RT}}[\frac{W}{\sqrt{1+\frac{2B}{v}+\frac{3C}{v^{2}}+...}}-\int{\frac{Bv^{2n-4}+3Cv^{2n-5}+...}{v^{\frac{n-1}{2}}[v^{n-1}+2Bv^{n-2}+3Cv^{n-3}+...]^{\frac{3}{2}}}W}dv]
\]
\par
\begin{equation}
-\sqrt{RT}[\int{\frac{(n-1)dv}{v\sqrt{1+\frac{2B}{v}+\frac{3C}{v^{2}}+...}}}+\int{\frac{(n-2)Bdv}{v^{2}\sqrt{1+\frac{2B}{v}+\frac{3C}{v^{2}}+...}}}+...]
\end{equation}
\par
\begin{equation}
=\sqrt{RT}[\int{\frac{dv}{v\sqrt{1+\frac{2B}{v}+\frac{3C}{v^{2}}+...}}}+\int{\frac{2Bdv}{v^{2}\sqrt{1+\frac{2B}{v}+\frac{3C}{v^{2}}+...}}}+\int{\frac{3Cdv}{v^{3}\sqrt{1+\frac{2B}{v}+\frac{3C}{v^{2}}+...}}}+...]
\end{equation}
\par
where W is work.
\end{theorem}
\par
\textbf{Proof}.\par Consider $(3.3)$,
\par
\begin{equation}
L^{T}=\int{\sqrt{\frac{RT}{v^{2}}+\frac{2RTB}{v^{3}}+\frac{3RTC}{v^{4}}+...+\frac{(n-1)RTY}{v^{n}}+\frac{nRTZ}{v^{n+1}}}}dv
\end{equation}
\par
It can be rewritten as
\par
\[
L^{T}=\int{\frac{\frac{RT}{v^{2}}+\frac{2RTB}{v^{3}}+\frac{3RTC}{v^{4}}+...+\frac{(n-1)RTY}{v^{n}}+\frac{nRTZ}{v^{n+1}}}{\sqrt{\frac{RT}{v^{2}}+\frac{2RTB}{v^{3}}+\frac{3RTC}{v^{4}}+...+\frac{(n-1)RTY}{v^{n}}+\frac{nRTZ}{v^{n+1}}}}}dv
\]
\par
\[
=\int{\frac{\frac{1}{v}[\frac{RT}{v}+\frac{RTB}{v^{2}}+\frac{RTC}{v^{3}}+...+\frac{RTY}{v^{n-1}}+\frac{RTZ}{v^{n}}]}{\sqrt{\frac{RT}{v^{2}}+\frac{2RTB}{v^{3}}+\frac{3RTC}{v^{4}}+...+\frac{(n-1)RTY}{v^{n}}+\frac{nRTZ}{v^{n+1}}}}}dv
\]
\par
\[
+\int{\frac{\frac{1}{v}[\frac{RTB}{v^{2}}+\frac{RTC}{v^{3}}+...+\frac{RTY}{v^{n-1}}+\frac{RTZ}{v^{n}}]}{{\sqrt{\frac{RT}{v^{2}}+\frac{2RTB}{v^{3}}+\frac{3RTC}{v^{4}}+...+\frac{(n-1)RTY}{v^{n}}+\frac{nRTZ}{v^{n+1}}}}}}dv
\]
\par
\[
+\int{\frac{\frac{1}{v}[\frac{RTC}{v^{3}}+...+\frac{RTY}{v^{n-1}}+\frac{RTZ}{v^{n}}]}{{\sqrt{\frac{RT}{v^{2}}+\frac{2RTB}{v^{3}}+\frac{3RTC}{v^{4}}+...+\frac{(n-1)RTY}{v^{n}}+\frac{nRTZ}{v^{n+1}}}}}}dv
\]
\par
\[
+...+\int{\frac{\frac{1}{v}[\frac{RTZ}{v^{n}}]}{{\sqrt{\frac{RT}{v^{2}}+\frac{2RTB}{v^{3}}+\frac{3RTC}{v^{4}}+...+\frac{(n-1)RTY}{v^{n}}+\frac{nRTZ}{v^{n+1}}}}}}dv
\]
\par
which gives
\par
\[
L^{T}=\int{\frac{pdv}{v\sqrt{\frac{RT}{v^{2}}+\frac{2RTB}{v^{3}}+\frac{3RTC}{v^{4}}+...+\frac{(n-1)RTY}{v^{n}}+\frac{nRTZ}{v^{n+1}}}}}
\]
\par
\[
+\int{\frac{(p-\frac{RT}{v})dv}{v\sqrt{\frac{RT}{v^{2}}+\frac{2RTB}{v^{3}}+\frac{3RTC}{v^{4}}+...+\frac{(n-1)RTY}{v^{n}}+\frac{nRTZ}{v^{n+1}}}}}
\]
\par
\[
+\int{\frac{(p-\frac{RT}{v}-\frac{RTB}{v^{2}})dv}{v\sqrt{\frac{RT}{v^{2}}+\frac{2RTB}{v^{3}}+\frac{3RTC}{v^{4}}+...+\frac{(n-1)RTY}{v^{n}}+\frac{nRTZ}{v^{n+1}}}}}
\]
\par
\[
+...+\int{\frac{(p-\frac{RT}{v}-\frac{RTB(T)}{v^{2}}-\frac{RTC(T)}{v^{3}}-...-\frac{RTY(T)}{v^{n-1}})}{v\sqrt{\frac{RT}{v^{2}}+\frac{2RTB}{v^{3}}+\frac{3RTC}{v^{4}}+...+\frac{(n-1)RTY}{v^{n}}+\frac{nRTZ}{v^{n+1}}}}}
\]
\par
Therefore, after rearranging, and considering that
\par
\[
\sqrt{\frac{RT}{v^{2}}+\frac{2RTB}{v^{3}}+\frac{3RTC}{v^{4}}+...+\frac{nRTZ}{v^{n+1}}}=\frac{\sqrt{RT}}{v}\sqrt{1+\frac{2B}{v}+\frac{3C}{v^{2}}+...+\frac{nZ}{v^{n-1}}}
\]
\par
we get
\par
\[
L^{T}=\frac{1}{\sqrt{RT}}[n\int{\frac{pdv}{\sqrt{1+\frac{2B}{v}+\frac{3C}{v^{2}}+...}}}-(n-1)\int{\frac{RTdv}{v\sqrt{1+\frac{2B}{v}+\frac{3C}{v^{2}}+...}}}
\]
\par
\begin{equation}
-(n-2)\int{\frac{RTBdv}{v^{2}\sqrt{1+\frac{2B}{v}+\frac{3C}{v^{2}}+...}}}-...]
\end{equation}
\par
where we drop the $n-th$ integral for simplicity. \par Considering
the first integral, we can integrate by parts considering
variables $\xi$ and $W$, (work), such that
\par
\[
\xi=\frac{1}{\sqrt{1+\frac{2B}{v}+\frac{3C}{v^{2}}+...+\frac{nZ}{v^{n-1}}}}\qquad
dW=pdv
\]
\par
and, since
\par
\[
\frac{d\xi}{dv}=\frac{Bv^{2n-4}+3Cv^{2n-5}+...}{v^{\frac{n-1}{2}}[v^{n-1}+2Bv^{n-2}+3Cv^{n-3}+...]^{\frac{3}{2}}}
\]
\par
then we have
\par
\begin{equation}
\int{\frac{pdv}{\sqrt{1+\frac{2B}{v}+\frac{3C}{v^{2}}+...}}}=\frac{W}{\sqrt{1+\frac{2B}{v}+\frac{3C}{v^{2}}+...}}-\int{\frac{Bv^{2n-4}+3Cv^{2n-5}+...}{v^{\frac{n-1}{2}}[v^{n-1}+2Bv^{n-2}+3Cv^{n-3}+...]^{\frac{3}{2}}}W}dv
\end{equation}
\par
Substituting $(3.9)$ into $(3.8)$, we get our final result
$(3.5)$. $(3.6)$ is immediate from $(3.4)$. \par
\begin{remark}
Note that while $(3.5)$ gives evidence that thermodynamic length
is \textit{also} work, $(3.6)$ is easier for computational
purposes.
\end{remark}
\par
Let's now look at specific cases. In particular, let's look at the
first, second and third expansion;i.e. $n=1$, $n=2$ and $n=3$.\par
For $n=1$, we have the Ideal (or quasi-Ideal) case.
\par
\begin{corollary}
Let
\par
\begin{equation}
p=\frac{RT}{v}
\end{equation}
\par
Then, along isotherms,
\par
\begin{equation}
L^{T}=\frac{1}{\sqrt{{RT}}}W=\sqrt{RT}\ln{(\frac{v_{2}}{v_{1}})}
\end{equation}
\par
where W is work given by $W=\int^{v_{2}}_{v_{1}}{p}dv$.
\end{corollary}
\par
\textbf{Proof I}.
\par
In this case, all the virial coefficients are zero and $n=1$. So,
from $(3.5)$ and $(3.6)$, we get $(3.11)$ immediately.
\par
During an isothermal process, the internal energy of an Ideal gas
remains constant because the gas temperature does not change.
Thus, $du=0$ which implies, by the first law of thermodynamics,
that if we do some work on a gas to compress it, the same amount
of energy will appear as heat transferred from the gas as it is
compressed. Thermodynamic length, in this case, seems to be a
measure of them up to a constant.
\par
\mathstrut
\par
Always along isotherms, it is easy to show that thermodynamic
length is work also in the case in which we consider the volume
occupied by molecules (quasi-ideal).\par In particular, if
$p=\frac{RT}{v-b}$ then $(3.11)$ still holds.
\par
\textbf{Proof II}.
\par
We look at the case in which
\par
\begin{equation}
(\frac{\partial^{2}{f}}{\partial{v}^{2}})_{T}=\frac{1}{RT}[(\frac{\partial{f}}{\partial{v}})_{T}]^{2}
\end{equation}
\par
where f is the molar Helmholtz potential. Naturally, $(3.12)$ is
equivalent to
\par
\begin{equation}
(\frac{\partial{p}}{\partial{v}})_{T}+\frac{1}{RT}p^{2}=0
\end{equation}
\par
since $(\frac{\partial{f}}{\partial{v}})_{T}=-p$. Now, $(3.13)$ is
a separable first order ordinary differential equation whose
solution is given by
\par
\[
p=\frac{RT}{v-b}
\]
\par
from which we have $f=-RT\ln{|v-b|}+h$, where h is any constant.
Then, by $(2.18)$
\par
\begin{equation}
L^{T}=\int{\sqrt{({\frac{\partial^{2}{f}}{\partial{v}^{2}}})_{T}}}dv=\int{\sqrt{{\frac{1}{RT}(\frac{\partial{f}}{\partial{v}})^{2}_{T}}}}dv=\int{\sqrt{{\frac{1}{RT}}}|(\frac{\partial{f}}{\partial{v}})_{T}|}dv=\sqrt{{\frac{1}{RT}}}\int|{p}|dv=\sqrt{{\frac{1}{RT}}}W
_{\blacksquare}
\end{equation}
\par
Let's consider, now, the case $n=2$ in which the only non-zero
virial coefficient is B.
\par
\begin{corollary}
Let
\par
\begin{equation}
p=\frac{RT}{v}+\frac{RTB}{v^{2}}
\end{equation}
\par
Then, along isotherms,
\par
\begin{equation}
L^{T}=\frac{1}{\sqrt{RT}}W+\sqrt{RT}[\ln(\frac{1+\frac{B}{v_{2}}+\sqrt{1+\frac{2B}{v_{2}}}}{1+\frac{B}{v_{1}}+\sqrt{1+\frac{2B}{v_{1}}}})-B(\frac{v_{2}-v_{1}}{v_{1}v_{2}})-2(\sqrt{1+\frac{2B}{v_{2}}}-\sqrt{1+\frac{2B}{v_{1}}})]
\end{equation}
\par
\begin{equation}
=2\sqrt{RT}[\ln(\frac{\sqrt{(v_{2}+2B)}+\sqrt{v_{2}}}{\sqrt{(v_{1}+2B)}+\sqrt{v_{1}}})-(\sqrt{1+\frac{2B}{v_{2}}}-\sqrt{1+\frac{2B}{v_{1}}})]
\end{equation}
\par
where the length is evaluated from volume $v_{1}$ to $v_{2}$ and
work is given by\par
$W=RT[\ln(\frac{v_{2}}{v_{1}})+B(\frac{v_{2}-v_{1}}{v_{1}v_{2}})]$.
\par
\end{corollary}
\par
\textbf{Proof}. From $(3.5)$ and $(3.6)$ after some calculation.
\par
Expression $(3.17)$ might be re-written in a more compact form by
setting $\rho_{i}=\sqrt{1+\frac{2B}{v_{i}}}$ with $i=1,2$. Then we
get
\par
\[
L^{T}=2\sqrt{RT}[\ln(\sqrt{\frac{v_{2}}{v_{1}}}(\frac{\rho_{2}+1}{\rho_{1}+1}))-(\rho_{2}-\rho_{1})]
\]
\par
Let's, now, denote by $W_{ideal}$, the work done on an Ideal gas
(see corollary $1$). It is interesting to note that, since
$W=RT[\ln(\frac{v_{2}}{v_{1}})+B(\frac{v_{2}-v_{1}}{v_{1}v_{2}})]=W_{ideal}+RTB(\frac{v_{2}-v_{1}}{v_{1}v_{2}})$,
then expression $(3.16)$ can be written as
\par
\[
L^{T}=\frac{1}{\sqrt{RT}}W_{ideal}+\sqrt{RT}[\ln(\frac{1+\frac{B}{v_{2}}+\sqrt{1+\frac{2B}{v_{2}}}}{1+\frac{B}{v_{1}}+\sqrt{1+\frac{2B}{v_{1}}}})-2(\sqrt{1+\frac{2B}{v_{2}}}-\sqrt{1+\frac{2B}{v_{1}}})]
\]
\par
\begin{equation}
=L^{T}_{ideal}+\sqrt{RT}[\ln(\frac{1+\frac{B}{v_{2}}+\sqrt{1+\frac{2B}{v_{2}}}}{1+\frac{B}{v_{1}}+\sqrt{1+\frac{2B}{v_{1}}}})-2(\sqrt{1+\frac{2B}{v_{2}}}-\sqrt{1+\frac{2B}{v_{1}}})]
\end{equation}
\par
It is evident that the second term on the right side of expression
$(3.18)$ gives the contribution to the thermodynamic length of an
isothermal quasi-static "real" process between two states due to
inter-particle interaction.
\par
\begin{remark}
\par
This result would help us to understand what thermodynamic length
is along isotherms for TD systems in which some interaction is
occurring. Note that, if $B=0$, like in the corollary $1$, then
length reduces to $(3.11)$.
\end{remark}
\par
\section{Appendix}
We include the case $n=3$ as a curiosity. We have the following
\par
\begin{corollary}
\par
Let
\par
\begin{equation}
p=\frac{RT}{v}+\frac{RTB}{v^{2}}+\frac{RTC}{v^{3}}
\end{equation}
\par
Then, along isotherms,
\par
\[
L^{T}=\sqrt{RT}[\ln{(\frac{\sqrt{v^{2}_{2}+2Bv_{2}+3C}+v_{2}+B}{\sqrt{v^{2}_{1}+2Bv_{1}+3C}+v_{1}+B})}+\frac{B}{\sqrt{3C}}[\ln{(\frac{\sqrt{3C}\sqrt{v^{2}_{2}+2Bv_{2}+3C}-Bv_{2}-3C}{\sqrt{3C}\sqrt{v^{2}_{1}+2Bv_{1}+3C}-Bv_{1}-3C})}
\]
\par
\begin{equation}
-\ln{(\frac{v_{2}}{v_{1}})}]-(\sqrt{1+\frac{2B}{v_{2}}+\frac{3C}{v^{2}_{2}}}-\sqrt{1+\frac{2B}{v_{1}}+\frac{3C}{v^{2}_{1}}})]
\end{equation}
\par
\end{corollary}
\par
\section{Conclusions}
\par
It would be interesting to see what thermodynamic length would be
along isotherms for different values of B and C or what the
physical meaning of length is, since, for $n=2$ above, work is
just a part of it. For example, the Van der Waals gas would be a
good starting point.
\par
\section{Acknowledgments}
\par
I gratefully thank Prof. S.Bleiler for reading my manuscript and
for replying with very useful comments. I also thank Prof. R.S.
Berry for his comments and many useful advices.
\par

\end{document}